\documentclass{PoS}

\title{Lepton Number Violation in Low Scale Seesaw Mechanism and its Collider Complementarity}

\ShortTitle{Lepton Number Violation in Low Scale Seesaw Mechanism}

\author{\speaker{Prativa Pritimita}, Nitali Dash, Sudhanwa Patra\\
       Center of Excellence in Theoretical and Mathematical Sciences,
Siksha 'O' Anusandhan University, Bhubaneswar-751030, India\\
       E-mail: \email{pratibha.pritimita@gmail.com}}

\abstract{We consider the TeV scale left-right symmetric theory which can accommodate low scale seesaw 
          mechanisms consistent with neutrino oscillation data and find new physics contributions to 
          neutrinoless double beta decay. The model facilitates natural type-II seesaw dominance and 
          the presence of extra particles make the Dirac neutrino mass matrix $M_D$  large that leads 
          to large light heavy neutrino mixing. The spontaneous symmetry breaking through doublets, 
          triplets and bidoublet scalars at TeV scale offers rich phenomenology accessible to LHC. 
          From the numerical studies of the new physics contributions to neutrinoless double beta decay 
          we derive a lower limit on absolute scale of lightest neutrino mass and find that normal hierarchy 
          (NH) pattern is favorable taking into account the cosmology and oscillation data. }

\FullConference{38th International Conference on High Energy Physics\\
		3-10 August 2016\\
		Chicago, USA}

\begin{document}

The origin of non-zero neutrino masses can be well-explained with low scale seesaw mechanisms that 
offer direct testability at LHC. Another low energy phenomenon that seems promising in the view that 
it can confirm Majorana nature of neutrinos is the neutrinoless double beta decay~\cite{Schechter:1981bd}. 
Interesting phenomenology can be expected when this process is studied in the framework of TeV scale left-right 
symmetric model (LRSM)~\cite{Mohapatra:1974gc,Senjanovic:1975rk}. We consider the extension of this model which 
comprises of usual quarks and leptons plus a sterile fermion per each generation and doublets, triplets, bidoublet 
scalars that help in the spontaneous symmetry breaking~\cite{Pritimita:2016fgr}. The gauge symmetry that governs the left-right symmetric model 
is $SU(2)_L \times SU(2)_R \times U(1)_{B-L} \times SU(3)_C$. When $SU(2)_R\times U(1)_{B-L}$ breaks at few TeV 
it leads to interesting collider signatures, neutrinoless double beta decay, lepton flavour violating processes 
and important implications to cosmology~\cite{Barry:2013xxa,Tello:2010am,Patra:2015bga, Deppisch:2015cua,
Ge:2015yqa,Deppisch:2014zta,Awasthi:2013ff,Chakrabortty:2012mh,Heeck:2015qra}. 
% The scalar sector comprises of 
% scalar doublets $H_{L,R}$, triplets $\Delta_{L,R}$ and a bidoublet $\Phi$ whereas fermion sector consists 
% of usual quarks $q_{L,R}$, leptons $\ell_{L,R}$ plus one extra sterile fermion $S_L$ per generation. 
The neutral lepton mass matrix in the flavor basis of $\left(\nu_L, S_L, N^c_R \right)$ is given by
\begin{eqnarray}
\mathbb{M} = \left(\begin{array}{c|ccc}   & \nu_L & S_L  & N^c_R   \\ \hline
\nu_L  & M_L       & 0       & M_D \\
S_L    & 0         & 0       & M \\
N^c_R  & M^T_D     & M^T     & M_R
\end{array}
\right).
\label{eq:numatrix-complete}
\end{eqnarray}
where $M_D$ is the Dirac neutrino mass connecting $\nu_L-N_R$, $M$ is the $N_R-S_L$ mixing 
matrix and $M_L (M_R)$ is the Majorana mass for LH (RH) neutrinos. Using the mass hierarchy 
$M_R > M > M_D \gg M_L$, the complete diagonalization procedure will lead to physical masses 
and mixing can be expressed in terms of light neutrino mass $m^{\rm d}_\nu=\mbox{diag}\{m_1,m_2,m_3\}$ 
and oscillation parameters as
\begin{eqnarray}
m_\nu = U_{\rm PMNS}\,\, m^{\rm d}_\nu\, \, U_{\rm PMNS}^T\, ,\quad 
M_N = \frac{v_R}{v_L} U_{\rm PMNS}\,\, m^{\rm d}_\nu\, \, U_{\rm PMNS}^T\, ,  
\quad M_S = -m^2_S~\left[\frac{v_R}{v_L} U_{\rm PMNS} m^{\rm d}_\nu U_{\rm PMNS}^T \right]^{-1}
\end{eqnarray}
The extra particles substantially make the Dirac neutrino mass matrix $M_D$ large that allows large 
light-heavy neutrino mixing. The charge current interaction Lagrangian with the physical masses and mixing for neutrinos is 
given by
\begin{eqnarray}
&&\hspace*{-1cm}\sum_{\alpha=e, \mu, \tau}
\bigg[\frac{g_L}{\sqrt{2}}\, \overline{\ell}_{\alpha \,L}\, \gamma_\mu {\nu}_{\alpha \,L}\, W^{\mu}_L 
      + \frac{g_R}{\sqrt{2}}\, \overline{\ell}_{\alpha \,R}\, \gamma_\mu {N}_{\alpha \,R}\, W^{\mu}_R \bigg] + \mbox{h.c.}
      \nonumber \\
&&\mbox{with}\quad \nu_\alpha=V^{\nu \nu}_{\alpha i} \nu_i + V^{\nu N}_{\alpha j} N_j + V^{\nu S}_{\alpha k} S_k    \nonumber \\ 
&&\quad ~ \quad ~ N_\beta = 0 \times \nu_i + V^{NN}_{\beta j} N_j  + V^{N S}_{\beta k} S_k   + \mbox{h.c.}        \nonumber 
\end{eqnarray}

\begin{center}
\begin{figure}
\hspace*{-0.5cm}
\includegraphics[width=0.55\linewidth]{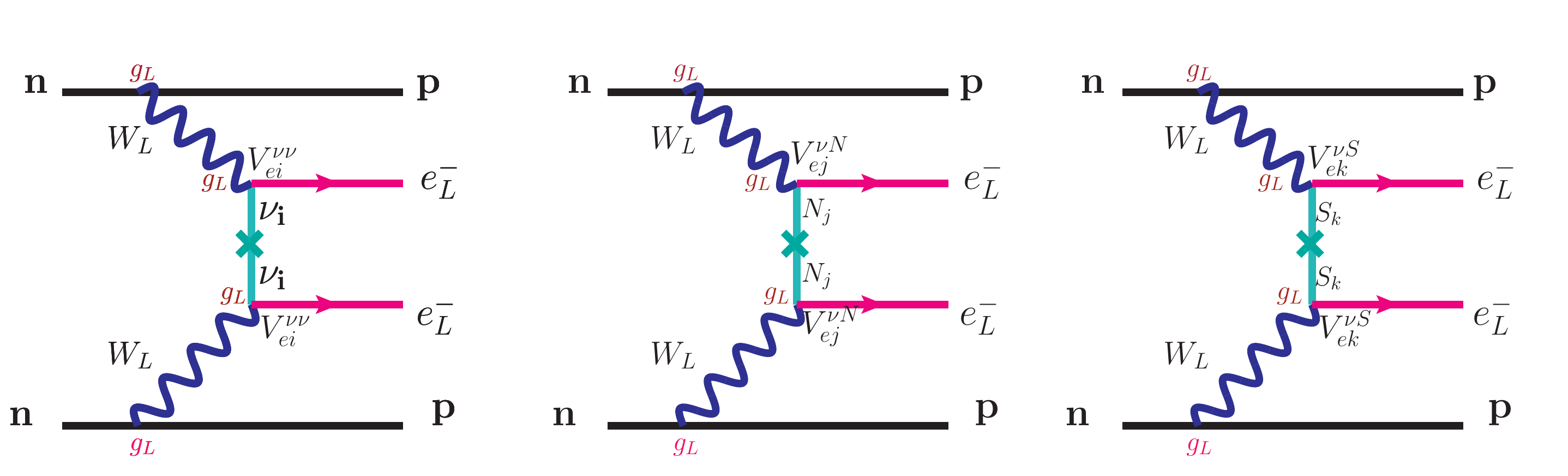}
\hspace*{0.2cm}
\includegraphics[width=0.55\linewidth]{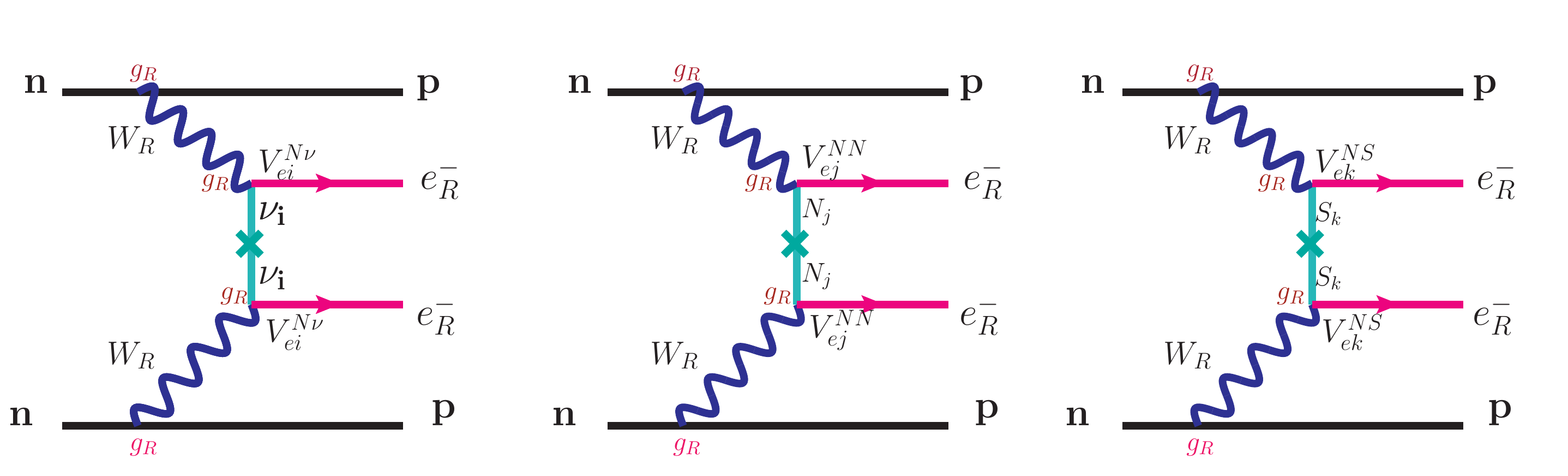}\\
\vspace*{0.9cm}
\hspace*{-0.5cm}
\includegraphics[width=0.55\linewidth]{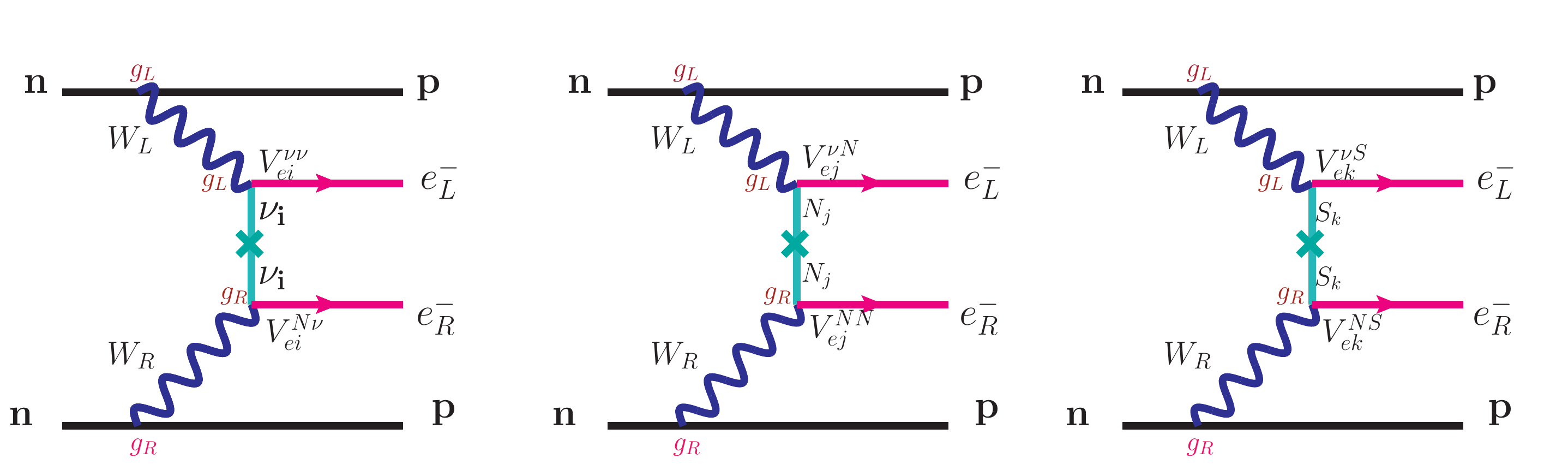}
\includegraphics[width=0.55\linewidth]{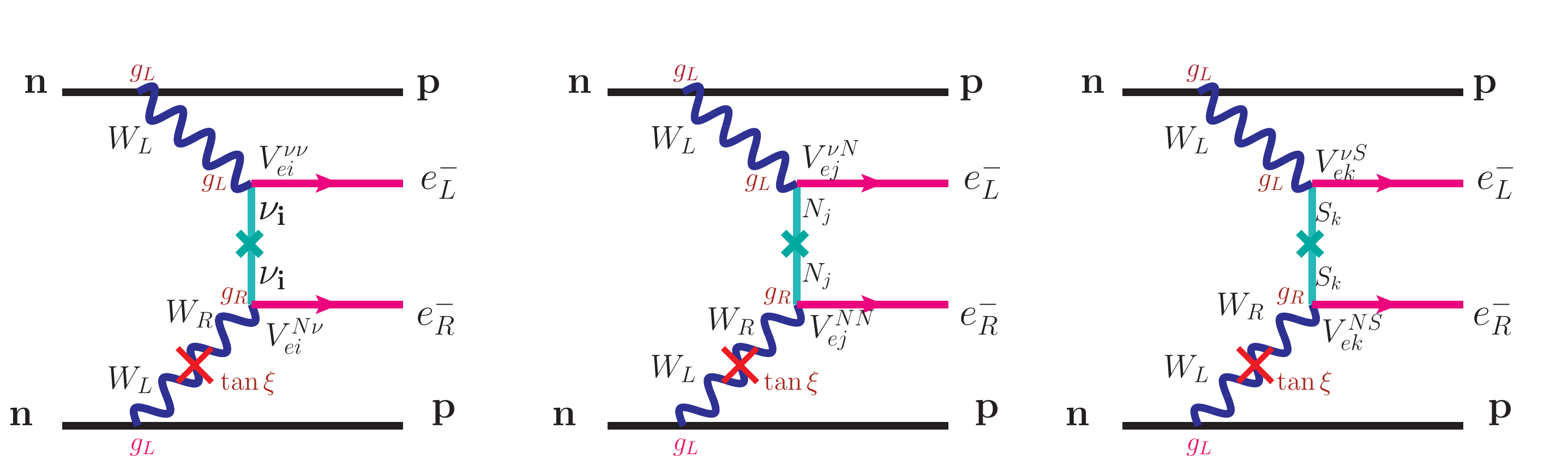}
\caption{Feynman diagrams for neutrinoless double beta decay with TeV spectrum of 
$W_R,Z_R$ gauge bosons, heavy neutrinos.}
\label{feyn}
\end{figure}
\end{center}

As an immediate result new contributions (displayed in Fig.\,\ref{feyn}) arise due to; 
\begin{center}
\begin{itemize}
\vspace*{-0.4cm}
 \item purely left-handed currents via exchange of heavy right-handed neutrinos and sterile neutrinos,
 \vspace*{-0.4cm}
 \item purely right-handed currents via exchange of heavy right-handed neutrinos and doubly charged scalar triplet, 
 \vspace*{-0.4cm}
 \item mixed helicity diagrams called as $\lambda$ diagrams involving light-heavy neutrino mixing.
 \item mixed helicity diagrams called as $\eta$ diagrams involving $W_L-W_R$ mixing.
\end{itemize}
\end{center}
 
\begin{center}
\begin{figure}
\includegraphics[width=0.450\textwidth]{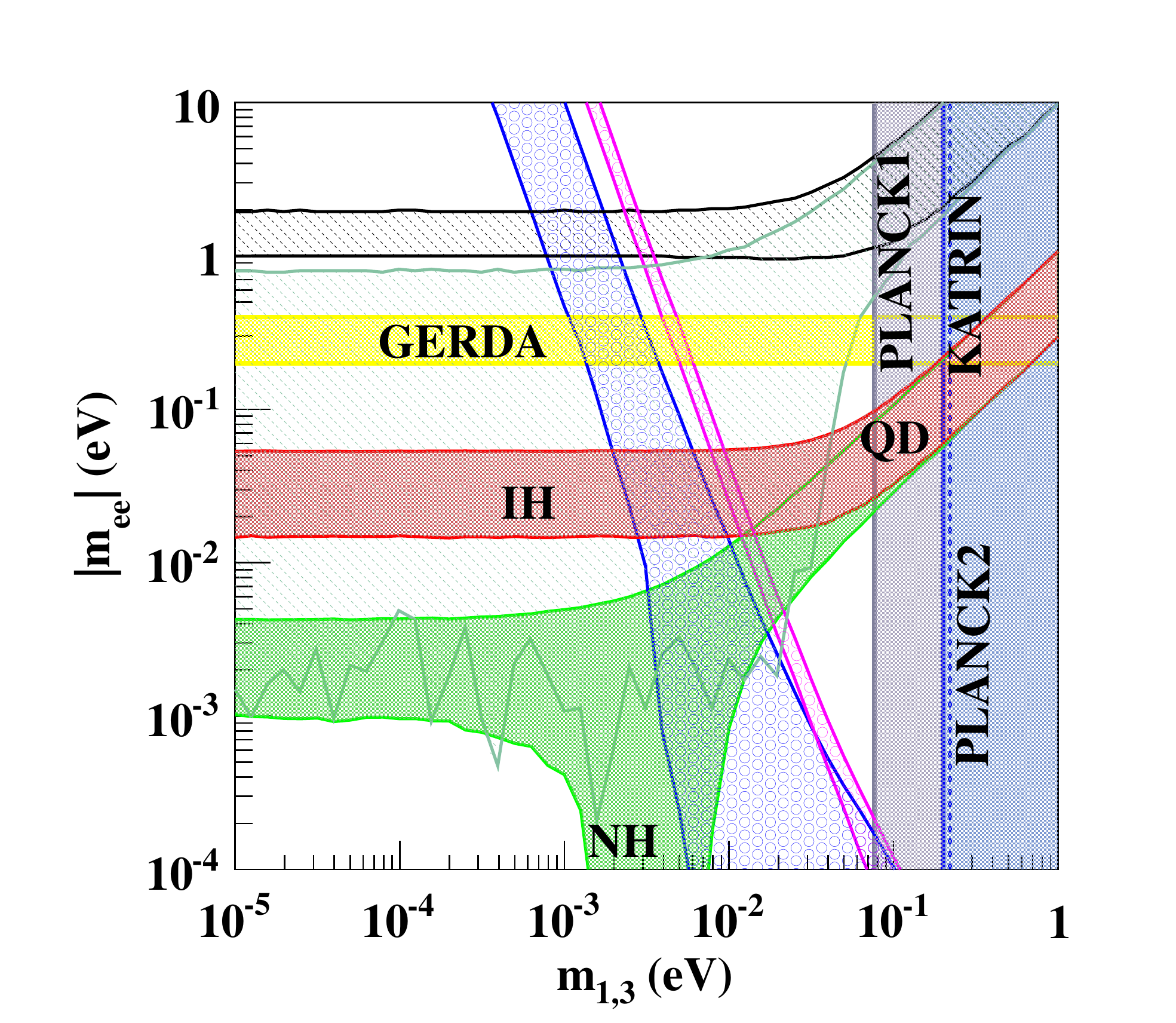}
\includegraphics[width=0.450\textwidth]{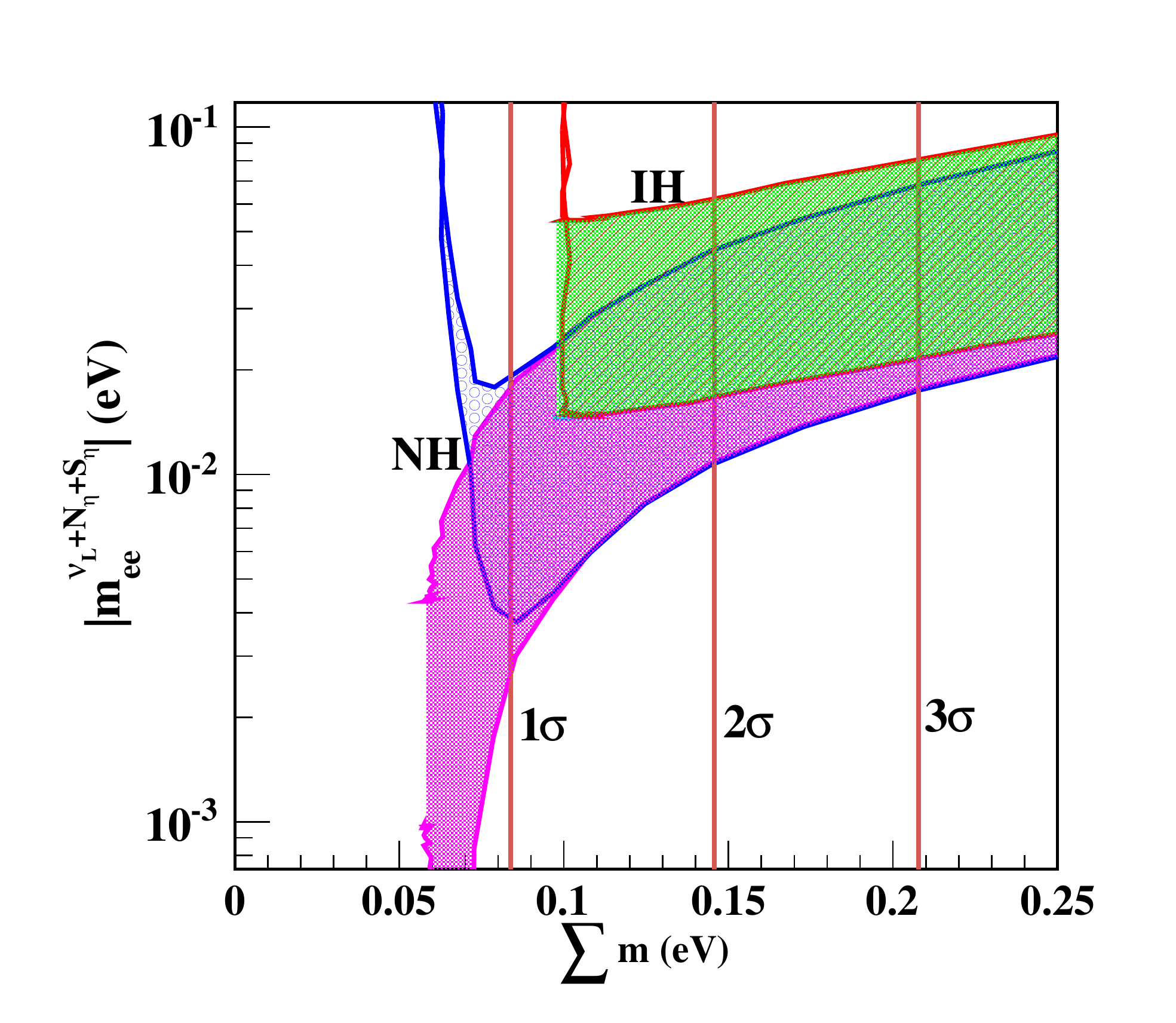}
\caption{New Physics contributions to neutrinoless double beta decay within TeV scale 
LRSM. The left-panel 
shows effective mass parameters vs lightest neutrino mass whereas the right-panel 
is for effective mass parameters vs sum of light neutrino masses. }
\label{plot}
\end{figure}
\end{center}

We minutely analyze these new physics contributions and derive the absolute scale 
of lightest neutrino mass using cosmology~\cite{Ade:2013zuv} and oscillation data~\cite{Gonzalez-Garcia:2014bfa}. 
The calculations looks simple since we have expressed all the physical masses and mixing in terms 
of lightest neutrino mass and neutrino oscillation parameters. We derive the lower limit on 
the absolute scale of light neutrino mass by saturating the experimental limit from GERDA~\cite{Agostini:2013mzu} and 
KamLAND-ZEN experiment~\cite{Gando:2012zm} as displayed in Fig.\,\ref{plot}. 

% \section{Conclusion}
We have discussed natural realization of type-II seesaw dominance within a 
      TeV scale LRSM. We have expressed all the physical masses and mixings like $\nu_L$, $N_R$ and $S_L$ in terms of oscillation 
      parameters and mass of lightest neutrino. We have derived the lower limit on absolute scale of lightest neutrino mass by numerically 
      estimating new physics contributions to $0\nu\beta\beta$ transition saturating 
      the current experimental limit. The effective mass vs sum of light neutrino masses shows that NH pattern of light neutrinos is favorable.

\end{document}